\documentclass[%
 aip,
 amsmath,amssymb,
 reprint,%
]{revtex4-1}

\usepackage[
    colorlinks=true,
    linkcolor=blue,
    citecolor=blue,
    urlcolor=blue]{hyperref}

\usepackage{graphicx}
\usepackage{dcolumn}
\usepackage{bm}

\usepackage[utf8]{inputenc}
\usepackage[T1]{fontenc}
\usepackage{mathptmx}
\usepackage{etoolbox}
\usepackage{xcolor}

\makeatletter
\def\@email#1#2{%
 \endgroup
 \patchcmd{\titleblock@produce}
  {\frontmatter@RRAPformat}
  {\frontmatter@RRAPformat{\produce@RRAP{*#1\href{mailto:#2}{#2}}}\frontmatter@RRAPformat}
  {}{}
}%
\makeatother
\begin{document}

\preprint{AIP/123-QED}

\title{Cs microcell optical reference at 459 nm with short-term frequency stability below 2 $\times$ 10$^{-13}$}
\author{E. Klinger}\email{emmanuel.klinger@femto-st.fr}
\altaffiliation{These authors contributed equally.}
\affiliation{Institut FEMTO-ST, Universit\'e Marie et Louis Pasteur, CNRS, SUPMICROTECH, Besan\c con, France}

\author{C. M. Rivera-Aguilar}
\altaffiliation{These authors contributed equally.}
\affiliation{Institut FEMTO-ST, Universit\'e Marie et Louis Pasteur, CNRS, SUPMICROTECH, Besan\c con, France}%

\author{A. Mursa}
\affiliation{Institut FEMTO-ST, Universit\'e Marie et Louis Pasteur, CNRS, SUPMICROTECH, Besan\c con, France}

\author{Q. Tanguy}
\affiliation{Institut FEMTO-ST, Universit\'e Marie et Louis Pasteur, CNRS, SUPMICROTECH, Besan\c con, France}

\author{N. Passilly}
\affiliation{Institut FEMTO-ST, Universit\'e Marie et Louis Pasteur, CNRS, SUPMICROTECH, Besan\c con, France}

\author{R. Boudot}
\affiliation{Institut FEMTO-ST, Universit\'e Marie et Louis Pasteur, CNRS, SUPMICROTECH, Besan\c con, France}

\date{\today}

\begin{abstract}
We describe the short-term frequency stability characterization of external-cavity diode lasers stabilized onto the $6\text{S}_{1/2}-7\text{P}_{1/2}$ transition of Cs atom at 459 nm, using a microfabricated vapor cell. The laser beatnote between two nearly-identical systems, each using saturated absorption spectroscopy in a simple retroreflected configuration, exhibits an instability of 2.5~$\times$~10$^{-13}$ at 1~s, consistent with phase noise analysis, and 3~$\times$~10$^{-14}$ at 200~s.  
The primary contributors to the stability budget at 1 second are the FM-AM noise conversion and the intermodulation effect, both emerging from laser frequency noise. 
These results highlight the potential of microcell-based optical references to achieve stability performances comparable to that of an active hydrogen maser in a remarkably simple architecture.
\end{abstract}

\maketitle
The optical interrogation of hot alkali atomic vapor within a microfabricated cell has led to the development of high-precision, integrated atomic instruments \cite{Kitching:APR:2018}. Pioneers among these devices, chip-scale microwave atomic clocks based on coherent population trapping 
have achieved remarkable development \cite{Lutwak:PTTI:2007,Yanagimachi:APL:2020,Carle:OE:2023}, including successful commercialization. These clocks, having a volume of 15--20~cm$^3$, a power consumption of 100--150~mW, and a timing error of a few microseconds per day, are now deployed in
navigation systems, instrumentation, metrology, and communications. Nevertheless, their short-term stability is limited by the laser frequency noise while their long-term stability is jeopardized by light-shifts and buffer-gas collisional shifts.

Over the last 20 years, a significant breakthrough in atomic clock performance has been achieved by transitioning from microwave to optical transitions \cite{Ye:2024}. However, this increase in performance has come in general with a dramatic increase of the complexity, size, and volume. Sub-Doppler spectroscopy (SDS) techniques \cite{Haroche}, that rely on the interaction of a hot atomic vapor confined in a cell with two counter-propagating laser fields, enable the detection of high Q-factor optical atomic resonances. These resonances are well-suited for laser frequency stabilization within a simple-architecture setup that does not require laser cooling. 

Optical references based on SDS techniques in cm-scale glass-blown hot vapor cells have demonstrated remarkable stability results. Most projects have involved the spectroscopy of the two-photon transition (TPT) of the Rb atom, at 778~nm \cite{Martin:PRAp:2018, Lemke:MDPI:2022, Li:OE:2024, Ruelle:EFTF:2024}, or using the 780 nm - 776~nm two-color scheme \cite{Perella:PRA:2019, Ahern:ArXiv:2024}. Other methods include modulation-transfer spectroscopy \cite{McCarron:IOP:2008} with iodine \cite{Doringshoff:PRAp:2017, Schuldt:GPS:2021, ClocksSea}, neutral ytterbium-174 \cite{Ahern:ArXiv:2024}, Rb \cite{Lee:OL:2023} and Cs atoms \cite{Miao:PRAp:2022}, or saturated absorption spectroscopy (SAS) in Cs \cite{Rovera_Cs_D2} and Rb \cite{Affolderbach:OLE:2005} atomic vapors. 

With the promising advancements in chip-scale lasers \cite{Corato:Nature:2023, Isichenko:SR:2024}, photonics, and microfabricated vapor cells \cite{Kitching:APL:2002, Douahi:EL:2007, Bopp:JPP:2021, Maurice:NMN:2022, Dyer:JAP:2022, Lucivero:OE:2022}, SDS techniques become attractive for the demonstration of fully-integrated optical references. For instance, in Ref. \cite{Hummon:Optica:2018}, a laser was stabilized to a Rb microcell using SAS, reaching a stability below 10$^{-11}$ from 1 to 10$^4$~s. The photonic integration of a microcell optical clock exploiting TPT of Rb atom at 778~nm was also reported \cite{Newman:Optica:2018}. Using an external-cavity diode laser (ECDL), it set a record for microcell optical references, achieving a short-term stability of 1.8~$\times$~10$^{-13}$ at 1~s and 2~$\times$~10$^{-14}$ after 100 s \cite{Newman:2021}. Competitive results were also reported with this approach on a transition of $^{87}$Rb atom \cite{Callejo:JOSAB:2025}. Dual-frequency sub-Doppler spectroscopy (DFSDS) \cite{MAH:OL:2016}, enabling the detection of high-contrast sign-reversed sub-Doppler resonances, has been used to stabilize an ECDL to a Cs microcell with a stability of 3~$\times$~10$^{-13}$ at 1~s and below 5~$\times$~10$^{-14}$ at 100~s  \cite{Gusching:OL:2023}. However, DFSDS requires the generation of a microwave-modulated optical field, usually obtained with an electro-optic modulator, which adds complexity to the system.

Although near-infrared transitions have been mainly used, recent advances in blue and near-ultraviolet integrated lasers \cite{Siddharth:APLPhot:2022}, combined with the two-fold increase of the reference transition frequency, constitute an interesting research path. In this domain, SDS of strontium at 461~nm was recently reported in a micromachined cell \cite{Pate:2023}. Nevertheless, strontium needs to be heated to high temperatures ($\sim$ 300$^{\circ}$C) to reach a sufficiently high vapor pressure, which poses challenges for maintaining the cell's lifetime. 
In Ref. \cite{Zhang:OL:2024}, an ECDL was locked to the $5\text{S}_{1/2} (F=2) \rightarrow 6\text{P}_{3/2} (F'=3)$ transition of Rb atoms in a tiny cubic cell, yielding a short-term stability of 2.2~$\times$~10$^{-12}$ at 1~s. Also, the exploration with SAS of the Cs atom $6\text{S}_{1/2} \rightarrow 7\text{P}_{1/2}$ transition at 459\;nm was recently initiated in a microfabricated cell \cite{Klinger:OL:2024}. However, no frequency stability characterization was reported in this work.

In the present paper, we report the short-term frequency stability characterization of ECDLs stabilized onto the Cs atom $6\text{S}_{1/2} \rightarrow 7\text{P}_{1/2}$ transition at 459~nm using SAS in a microfabricated vapor cell. In comparison with Ref. \cite{Klinger:OL:2024}, where separated pump and probe beams were used, we use here SAS in its simplest retro-reflected configuration. Two laser systems were assembled such that a beatnote, obtained by shifting the frequency of one laser with an acousto-optic modulator (AOM), can be analyzed and counted. The Allan deviation of the laser beatnote, when both lasers are locked, is lower than 2.5~$\times$~10$^{-13}$ at 1 s and 3~$\times$~10$^{-14}$ at 200~s. The stability at 1 s of one single laser is then estimated to be 1.8~$\times$~10$^{-13}$, a level competitive with best stability results reported so far for MEMS cell-based optical references. The primary factors affecting the short-term stability originate from the laser FM noise, through the FM-AM conversion process and the intermodulation effect \cite{Audoin:TIM:1991}. This suggests that there is room for improvement using lasers with lower FM noise. 

\begin{figure}[t!]
    \centering
    \includegraphics[width=0.95\linewidth]{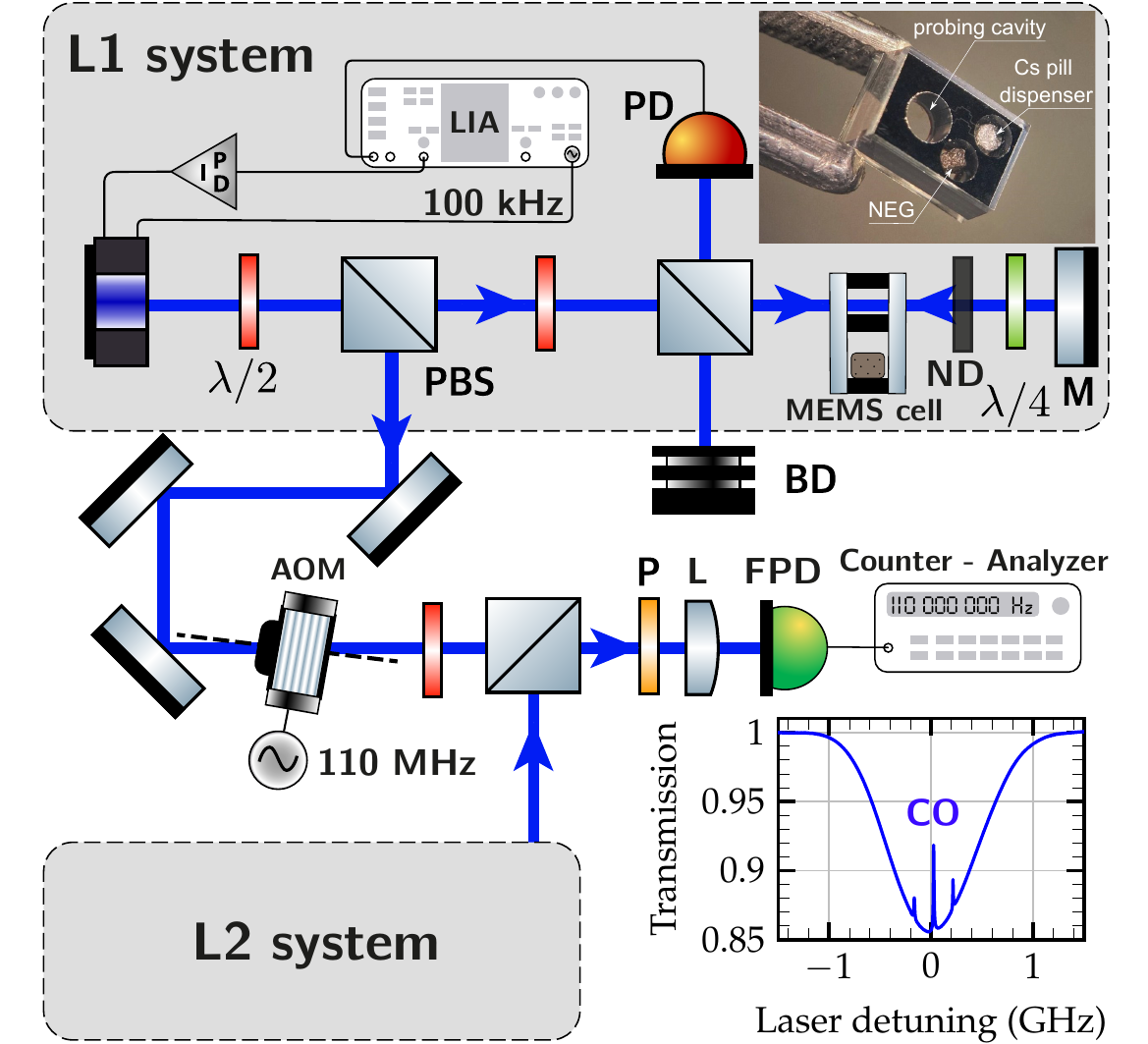}
    \caption{Experimental setup. Two ECDLs, emitting at 459~nm, are stabilized onto a microfabricated Cs vapor cell, shown in the bottom-right corner, using SAS. The frequency of the laser system 1 (L1) is frequency-shifted by 110 MHz using an acousto-optic modulator (AOM). A laser beatnote between the two lasers, detected at 110~MHz with a fast photodiode (FPD), is filtered, amplified and analyzed. PBS: polarizing beam splitter, BD: beam dump, ND: neutral density filter, M: mirror, PD: photodiode, LIA: lock-in amplifier, PID: PID controller, P: polarizer, L: converging lens.}
    \label{fig:1}
\end{figure}

Figure \ref{fig:1} shows the experimental setup. Two ECDLs (L1 and L2, Toptica DL-Pro), emitting at 459~nm, are each stabilized onto a Cs vapor microfabricated cell using SAS. An optical isolation stage of 35~dB is placed at the output of the laser head to prevent feedback. The linearly-polarized laser beam crosses a half-wave plate, a polarizing beam splitter (PBS), and is transmitted, as the pump beam ($P_L\simeq14.3\;$mW), through the Cs vapor cell. At the output of the cell are placed a neutral density filter (ND), a quarter-wave plate and a mirror that provides the counter-propagating beam, orthogonally polarized to the incident one, and used as the probe beam ($P_r$~$\simeq$~500~$\mu$W). The reflected beam is then directed by the PBS towards a photodiode (PD) that delivers the spectroscopic signal. The inset of Fig.\;\ref{fig:1} shows the sub-Doppler resonances detected at the bottom of the Doppler-broadened profile. Due to its higher amplitude, we chose to lock the lasers onto the crossover (CO) line, involved between the $F = 4 \rightarrow 3'$ and $F = 4 \rightarrow 4'$ transitions, separated by 377.6(2)\;MHz \cite{Williams:LPL:2018}. For laser frequency stabilization, a dispersive zero-crossing error signal is generated from the atomic signal by modulating the laser current (modulation frequency $f_m = 100\;$kHz), and applying lock-in detection. The error signal is fed into a digital proportional-integral controller whose output is used to correct the laser current. 

\begin{figure}[t!]
    \centering
    \includegraphics[width=0.9\linewidth]{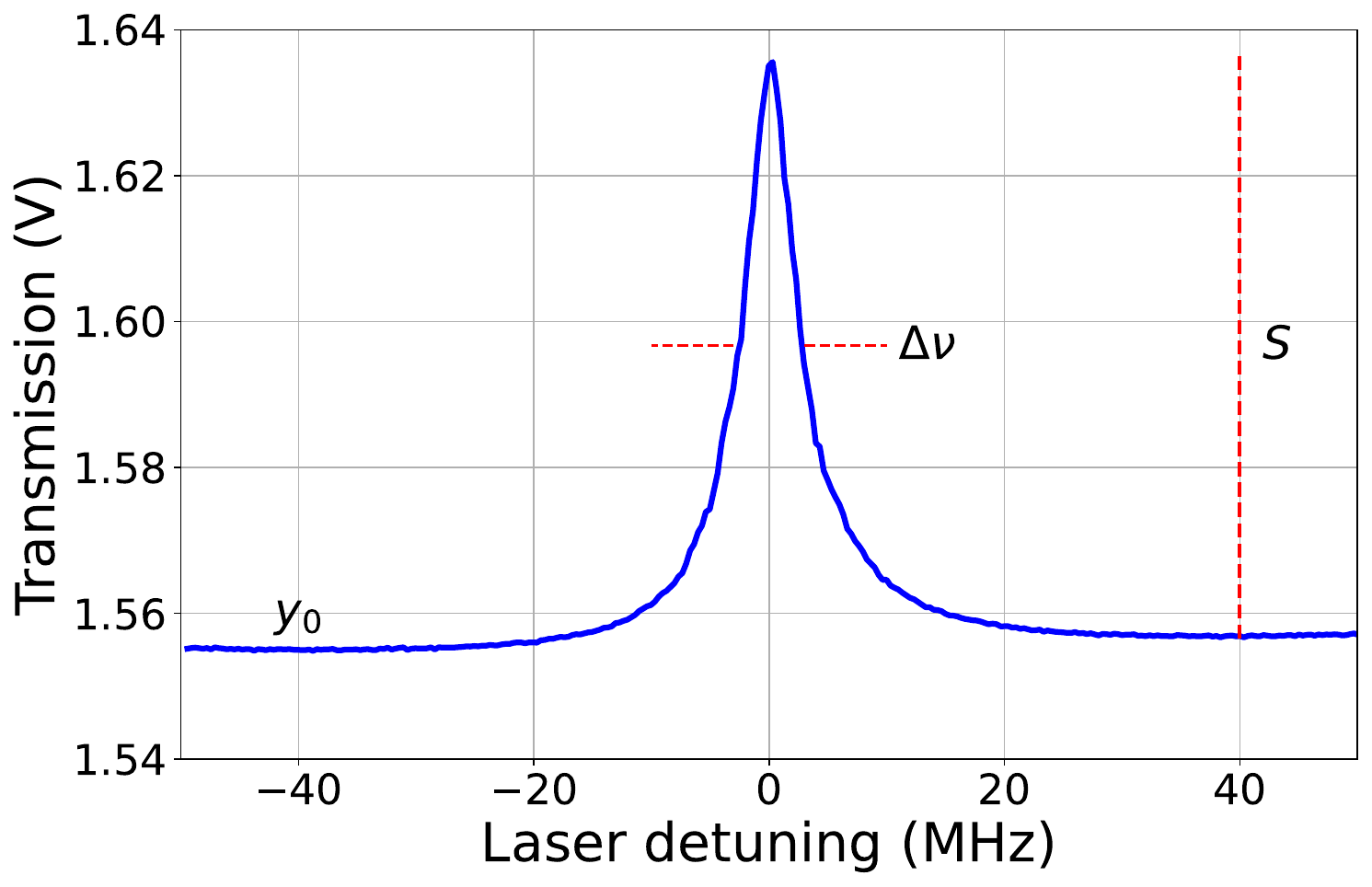}
    \caption{Central cross-over sub-Doppler resonance used for laser frequency stabilization.}
    \label{fig:2}
\end{figure}

The cell technology is adapted from the one described in Ref. \cite{Klinger:OL:2024}. It involves etching two adjacent cavities in silicon, and enclosing them between two anodically-bonded glass wafers. 
The first cavity contains a pill dispenser, 
which is laser-activated after the final sealing in order to fill the cell with Cs vapor \cite{Douahi:EL:2007, Vicarini:SA:2018}. The second cavity, in which atom-light interaction takes place, is 2-mm in diameter and 1.5-mm long, and connected by thin channels to the first cavity. In Ref. \cite{Klinger:OL:2024}, the zero-power linewidth of the sub-Doppler resonance was measured to be about $ 8\;$MHz, which is significantly broader than the transition natural linewidth ($\sim 950\;$kHz), with the increase attributed to collisional broadening \cite{Pitz:PRA:2009}. In the present work, aiming to enhance the cell purity, a third cavity has been incorporated into the cell preform to host a passive non-evaporable getter (NEG) \cite{Boudot:SR:2020}, which can also be activated by laser. Additionally, alumino-silicate glass (ASG) is employed for the cell windows in order to minimize helium permeation \cite{Dellis:2016, Carle:JAP:2023}. The MEMS cell, shown in Fig.\;\ref{fig:1}, is placed within a physics package, controlled at a temperature set-point of about 118$^{\circ}$C and covered by a single-layer mu-metal magnetic shield.

\begin{figure}[t!]
    \centering
    \includegraphics[width=0.85\linewidth]{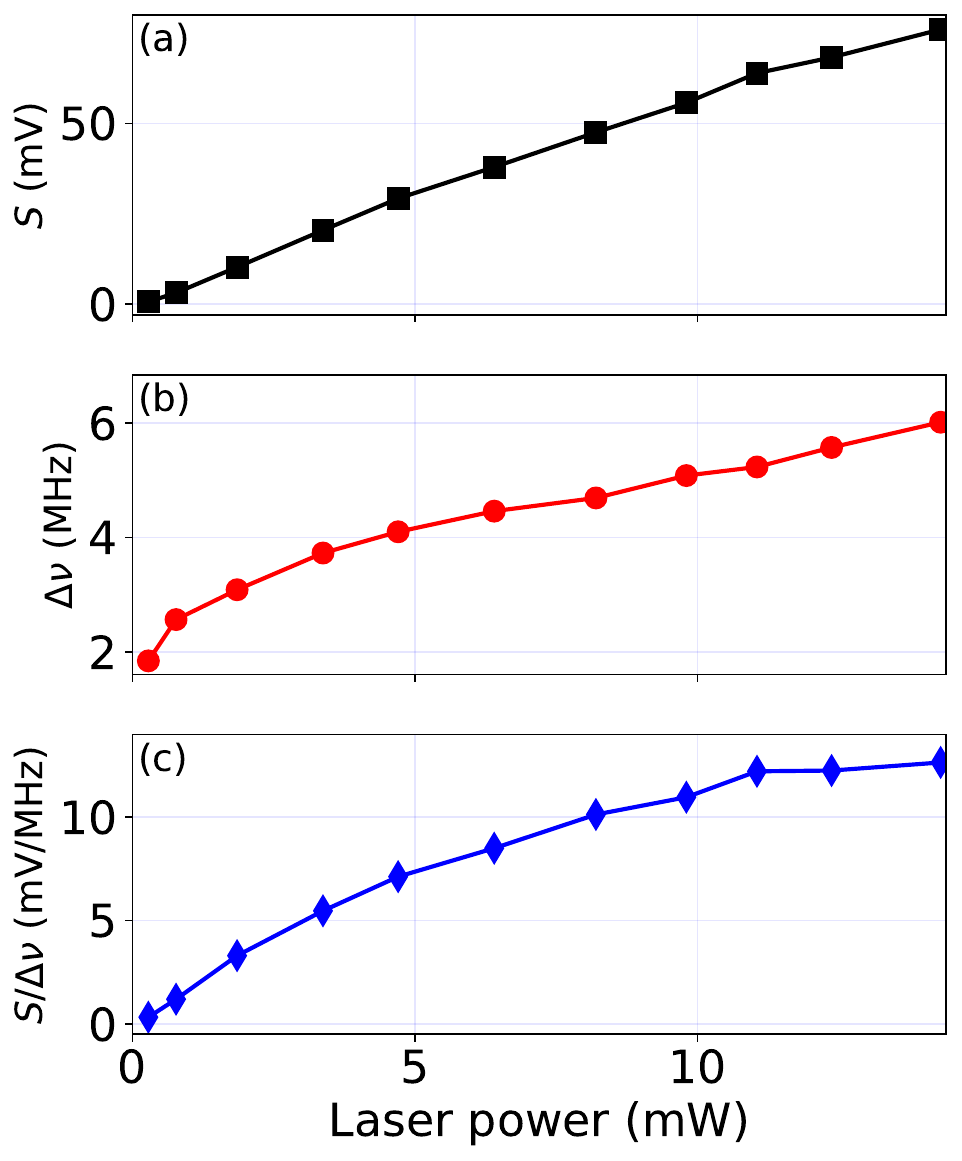}
    \caption{Signal $S$ (a), linewidth $\Delta \nu$ (b) and signal/linewidth ratio (c) of the sub-Doppler resonance versus the total laser power $P_L$ at the cell input, for laser 1. }
    \label{fig:3}
\end{figure}

For stability characterization, a laser beatnote, obtained between both laser systems by shifting the frequency of the laser~1 with an AOM driven at 110 MHz, was detected, filtered, amplified and sent to a frequency counter (HP53132A) or phase noise analyzer (RS FSWP). The frequency counter was referenced to an active hydrogen maser.

Figure \ref{fig:2} provides a close-up view of the central CO resonance. Fitting the resonance by a Lorentzian function yields a resonance linewidth $\Delta \nu \simeq 6\;$MHz, a signal height $S\simeq 76\;$mV, a sensitivity $S_l = S / \Delta \nu \simeq$ 12.6 mV/MHz, and a contrast $C = S/y_0$, with $y_0$ the dc background of the resonance line, of 4.8\;\%. Figure \ref{fig:3} shows the evolution of the sub-Doppler resonance signal $S$, linewidth $\Delta \nu$ and ratio $S / \Delta \nu$ versus the total laser power $P_L$ at the cell input, for laser 1. In these tests, the pump power to probe power ratio $r = P_L/P_r$ is about 27.4. The increase of the pump beam power yields an increase of the signal $S$, at the expense of power-broadening of the resonance. At the smallest tested laser power ($P_L \sim 0.28\;$ mW), we find $\Delta \nu \simeq 1.8\;$MHz. This value is significantly smaller than the linewidths reported in \cite{Klinger:OL:2024},  indicating an improved purity of the cell inner atmosphere. The signal/linewidth ratio increases with the laser power before reaching a plateau at about 15\;mW. Similar features were obtained for the laser 2. We chose to operate both laser systems with an incident power $P_L$~$\simeq 14.3\;$mW.

\begin{figure}[t!]
    \centering
    \includegraphics[width=0.90\linewidth]{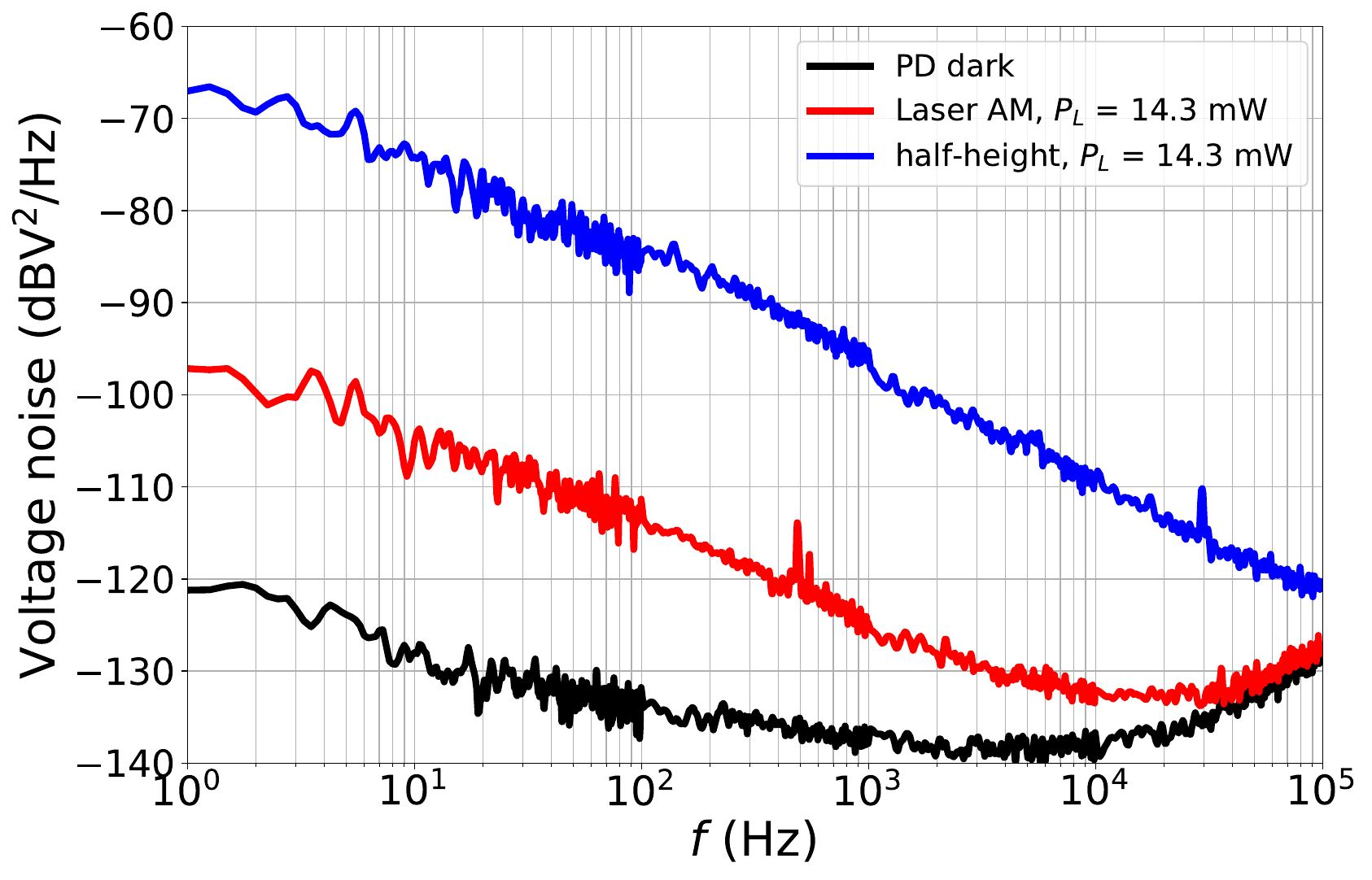}
    \caption{Detection noise at the photodiode output in different conditions: photodiode in the dark, laser frequency tuned out of the Doppler-broadened optical resonance (laser AM noise), half-height of the sub-Doppler resonance.}
    \label{fig:4}
\end{figure}

Following this spectroscopic study, we have measured the total detection noise at the photodiode (PD) output in different conditions. Results are shown in Fig.\;\ref{fig:4}. The first curve (black) shows the voltage noise of the PD "in the dark" (laser beam blocked). The second curve (red), image of the laser amplitude noise (AM), was measured by tuning the laser frequency out of the Doppler-broadened optical resonance. At $f=f_m= 100\;$kHz, the laser AM noise is as low as the photodetector noise, at the level of $-130\;$dBV$^2$/Hz. The last curve (blue) shows the total detection noise measured at half-height of the sub-Doppler resonance. In this case, we observe a clear increase of the noise level, reaching at $f=100\;$kHz the level of $-120\;$dBV$^2$/Hz (i.e. $1\;\mu$V/$\sqrt{\rm{Hz}}$). With the data obtained in Figs.\;\ref{fig:2} and \ref{fig:4}, the signal-to-noise ratio in a 1\;Hz bandwidth of the detected resonance is found to be $7.6 \times 10^4$.

\begin{figure}[t!]
    \centering    \includegraphics[width=0.90\linewidth]{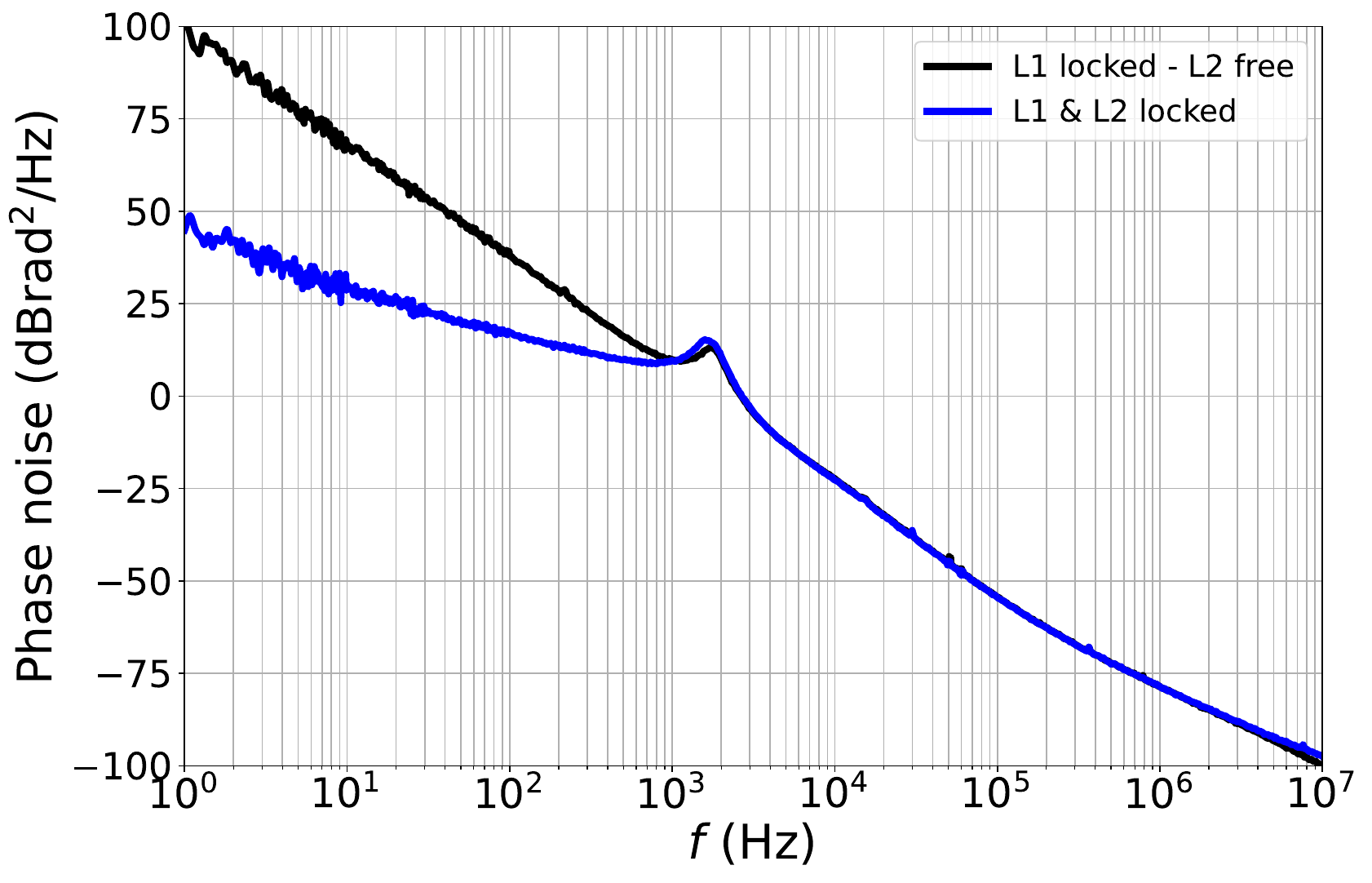}
    \caption{Phase noise of the laser beatnote in free-running (L1 locked and L2 free) and locked conditions.}
    \label{fig:5}
\end{figure}

Figure \ref{fig:5} shows the phase noise of the laser beatnote, in two conditions. In the free-running case (L1 locked and L2 free), the phase noise of the laser beatnote is $-54\;$dBrad$^2$/Hz, $-63\;$dBrad$^2$/Hz and $-78.5\;$dBrad$^2$/Hz at $f=100\;$kHz, 200\;kHz and 1\;MHz, respectively. Comparable results were obtained in the case where L1 is free and L2 is locked. We assume that the phase noise of a single laser is 3 dB lower than the beatnote noise ($-$66~dBrad$^2$/Hz at $f=2 f_m =200\;$kHz). The contribution $\sigma_{int}$ of the intermodulation effect \cite{Audoin:TIM:1991} on the atomic microcell reference stability at 1\;s is 
   $\sigma_{int}\simeq\frac{f_m}{\nu_0} \sqrt{S_{\varphi}(2f_m)}$, 
where $\nu_0$ is the laser frequency ($\nu_0 \simeq 6.5 \times 10^{14}\;$Hz) and $S_{\varphi}(2f_m)$ is the
phase fluctuations power spectral density
of the laser at $f = 2 f_m$. Here, with $f_m = 100\;$kHz, we estimate
$\sigma_{int} \simeq 7.8~\times~10^{-14}$ for a single laser. At $f$~=~1 Hz, the phase noise spectrum of the laser beatnote exhibits a $f^{-3}$ slope, signature of flicker frequency noise, with $b_{-3}=+100\;$dBrad$^2$/Hz ($+$~97 dBrad$^2$/Hz for a single laser), from which a stability
\begin{equation}
\sigma_{y}(1\,\text{s})=2\ln 2 \sqrt{(b_{-3}/\nu_0^2)}=2.2\times 10^{-10}
\end{equation}
 can be extracted \cite{Rubiola_phase_noise_oscillators}. In the locked case, the phase noise spectrum of the laser beatnote shows a $f^{-2}$ trend at 1\;Hz, signature of white frequency noise, with, $b_{-2}=+45$\;dBrad$^2$/Hz ($+42\;$dBrad$^2$/Hz for a single laser), from which a stability 
 \begin{equation}
\sigma_{y}(1\,\text{s})=\sqrt{(b_{-2}/\nu_0^2)}=1.9\times 10^{-13}
 \end{equation}
 is expected for a single laser. The bandwidth of the laser frequency servos is about 1.5~kHz in both systems, as revealed by the bumps on the noise spectra.

\begin{figure}[t!]
    \centering    \includegraphics[width=0.9\linewidth]{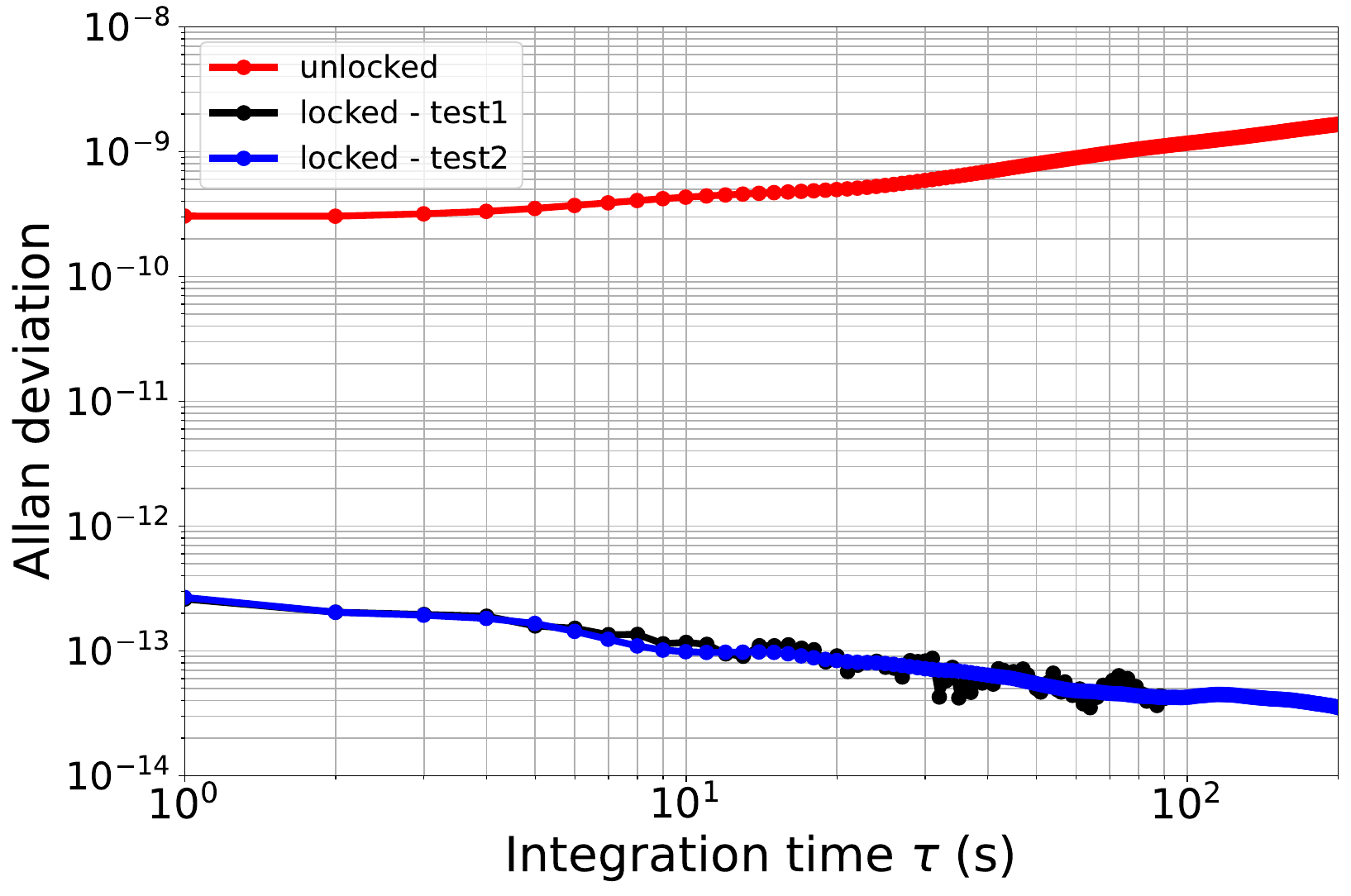}
    \caption{Allan deviation of the laser beatnote in free-running and locked conditions. Two tests were done in locked conditions giving similar results.}
    \label{fig:6}
\end{figure}

Figure\;\ref{fig:6} shows the Allan deviation of the laser beatnote, in free-running and locked conditions. In the free-running case, the fractional frequency stability of the laser beatnote is about $3\times10^{-10}$ at 1\;s, in reasonable agreement with the phase noise analysis,
and reaches 10$^{-9}$ at 100~s. In locked conditions, the Allan deviation of the laser beatnote is $2.5\times10^{-13}$ at 1\;s. Assuming that both lasers contribute equally, the stability of a single laser is estimated to be 1.8~$\times$~10$^{-13}$ at 1~s. These short-term performances rival the best microcell-based optical references. \cite{Newman:2021, Gusching:OL:2023}. The Allan deviation plot shows a plateau, possibly caused by the lasers' frequency sensitivity to cell temperature, before averaging down to $3\times 10^{-14}$ at 200~s.

The stability budget, reported in Table\;\ref{tab:budget} and established using the expressions found in \cite{Gusching:JOSAB:2021}, is in good agreement with the measured stability.
The main contribution to the stability at 1\;s is currently the FM-AM conversion process, followed by the intermodulation effect. This suggests that the primary factor influencing stability at 1~s is the FM-AM conversion process, followed by the intermodulation effect \cite{Audoin:TIM:1991}. Additional contributions include the laser AM noise and photodetector noise, currently estimated to be in the mid-10$^{-14}$ range.  

\begin{table}[t]
\caption{Short-term stability budget of microcell-stabilized laser.  The laser power at the photodiode input, used for photon shot noise contribution, is 460~$\mu$W.}
\label{tab:budget}\vspace*{-0.05ex} 
\begin{center}\begin{tabular}{ll}\hline
Noise source  & $\sigma$ (1 s) \\ \hline
Laser FM-AM noise & 1.2 $\times$ 10$^{-13}$ \\
Intermodulation & 7.8 $\times$ 10$^{-14}$ \\
Laser AM noise & 3.8 $\times$ 10$^{-14}$ \\
PD dark & 3.8~$\times$~10$^{-14}$ \\
Photon shot noise & 8.3~$\times$~10$^{-15}$ \\
\hline
$\sigma_{y}$ (1 s) - single laser & 1.6 $\times$ 10$^{-13}$ \\
$\sigma_{y}$ (1 s) - laser beatnote & 2.3 $\times$ 10$^{-13}$ \\
\hline
\end{tabular}\end{center}
\end{table}

In conclusion, we have reported the short-term stability characterization of ECDLs stabilized onto microfabricated cells at 459 nm using SAS of the Cs atom 6S$_{1/2}$-7P$_{1/2}$ transition. The Allan deviation of the laser beatnote between two nearly-identical systems was measured to be 2.5~$\times$~10$^{-13}$ at 1~s and 3~$\times$~10$^{-14}$ at 200~s. These performances rival the best stability results reported so far for an optical reference based on a microfabricated cell. The two main contributions to the short-term stability originate from the laser FM noise, through the FM-AM conversion process and the intermodulation effect. This suggests room for improvement with the use of lasers exhibiting lower FM noise \cite{kervazo2024sub}. 

This work was supported by Centre National d'Etudes Spatiales (CNES), in the frame of the OSCAR project, Agence Nationale de la Recherche (ANR) through LabeX FIRST-TF (Grant ANR 10-LABX-48-01) (LEILA project), and EIPHI Graduate school (Grant ANR-17-EURE-0002) (REMICS project). The PhD thesis of C. Rivera-Aguilar is co-funded by the program FRANCE2030 QuanTEdu (Grant ANR-22-CMAS-0001) and CNES. This work was supported by the French RENATECH network through its FEMTO-ST technological facility (MIMENTO), and by the Oscillator-IMP platform.

\section*{Data availability statement}
The data supporting the findings of this study are available from the corresponding author upon reasonable request.

\section*{Conflict of interest}
The authors state that there is no conflict of interest to disclose.\\



\bibliography{BiblioCloks}

\end{document}